\documentclass[aps,prl,showpacs,twocolumn]{revtex4}

\usepackage{amssymb}
\usepackage{epsfig}
\usepackage{graphicx}
\usepackage{amsmath}
\usepackage{array,color}

\begin{document}

\title{Effects of Geometrical Symmetry on the Vortex
Nucleation and Penetration in Mesoscopic Superconductors}
\author{Xing-Hua Hu$^1$, An-Chun Ji$^2$, Xiang-Gang Qiu$^{1}$}
\email{xgqiu@aphy.iphy.ac.cn}\author{Wu-Ming Liu$^1$}
\address{$^1$Beijing National Laboratory for Condensed Matter Physics,
Institute of Physics, Chinese Academy of Sciences, Beijing 100190,
China}
\address{$^2$Center of Theoretical Physics, Department of Physics, Capital Normal University, Beijing 100048, China}
\date{\today}
\begin{abstract}
We investigate how the geometrical symmetry affects the penetration
and arrangement of vortices in mesoscopic superconductors using
self-consistent Bogoliubov-de Gennes equations. We find that the
entrance of the vortex happens when the current density at the hot
spots reaches the depairing current density. Through determining the
spatial distribution of hot spots, the geometrical symmetry of the
superconducting sample influences the nucleation and entrance of
vortices. Our results propose one possible experimental approach to
control and manipulate the quantum states of mesoscopic
superconductors with their topological geometries, and they can be
easily generalized to the confined superfluids and Bose-Einstein
condensates.
\end{abstract}


\maketitle

The advances in modern nanotechnology and the development of quantum
computing have opened many new perspectives for research on
mesoscopic superconductors \cite{moshchalkov}. One fascinating
aspect in mesoscopic superconductivity is the novel physics
associated with vortices, which has been a subject of great
experimental
\cite{victor,hess2,chibotaru,grigorieva,crabtree,nishio2,cren,geim1,geim2}
and theoretical \cite{brandt,victorprb,baelus,bonca} interest in the
past decades. In the mesoscopic superconductors, vortices are
quantized and confined by the  sample geometry, so they can exhibit
many unique phenomena compared with conventional bulk
superconductors. For example, when the sample is mesospcopic, the
giant vortex can form, and there may exhibit the exotic paramagnetic
Meissner effect \cite{victorprb}. Further, it is shown that the
symmetry of the sample geometry can dramatically affect the
properties of the mesoscopic system \cite{victor}. Recent experiment
has also reported the symmetry-induced antivortices formation
\cite{chibotaru}, which is a unique character of a mesoscopic
superconductor. However, how the geometrical symmetry of sample can
control the vortex state is hitherto not well understood.

In this Letter, we explore the effect of symmetry on the vortex
nucleation and entrance in the mesoscoopic superconductor far below
the critical temperature. We develop an effective numerical method
to generally solve the Bogoliubov-de Gennes (BdG) equations
\cite{gennes}, which is based on the finite element method (FEM)
\cite{zienkiewicz, suematsu}.  Compared with the conventional
Ginzburg-Landau (GL) theory \cite{tinkham}, the BdG equations work
well in wider temperature region, and can give the spectrum of
excitations for spatially inhomogeneous superconductor
self-consistently, supposed the few fundamental material parameters
are given.  Several groups have successfully used the BdG equations
to study the single vortex line \cite{hayashi,kato,melnikov,gygi}.

Here, we solve the BdG equations self-consistently, and obtain the
vortex pattern and the current density in mesoscopic superconductors
with arbitrary and complicated geometries. Our results show that the
current distribution and the penetration of vortices are determined
by the symmetry of the sample geometry, and the entrance of the
vortex happens only when the current density at the hot spots (i.e.
the spots with maximum current density) reaches the depairing
current density. These facts reveal unambiguously that the
geometrical symmetry of the superconducting sample influences the
nucleation and entrance of vortices, {\it through determining the
spatial distribution of hot spots}. These results provide a
practicable route to manipulate the quantum states of a mesoscopic
superconductors in future applications.

We start with the BdG equations for the quasiparticle wave functions
$u_n(\textbf{r})$ and $v_n(\textbf{r})$ in the presence of a
magnetic field

\begin{eqnarray}
&&\left[\frac{1}{2m}(\frac{\hbar}{i}\nabla-\frac{e\textbf{A}}{c})^2
-\mu\right]u_n(\textbf{r})+\triangle(\textbf{r})
v_n(\textbf{r})=E_n
u_n(\textbf{r}),\nonumber\\
&&\left[\frac{-1}{2m}(\frac{\hbar}{i}\nabla+\frac{e\textbf{A}}{c})^2
+\mu\right]v_n(\textbf{r})+\triangle^{*}(\textbf{r})u_n(\textbf{r})=E_n
v_n(\textbf{r}),\nonumber\\
\label{BdG}
\end{eqnarray}
where $E_n$ is the $n$-th energy eigenvalue, $\triangle(\textbf{r})$
the pair potential, $\textbf{A}(\textbf{r})$ the vector potential
and $\mu$ the chemical potential.

The pair potential $\triangle(\textbf{r})$ is determined
self-consistently by
\begin{equation}
 \triangle(\textbf{r})=g\sum^n_{|E_n|\leq E_c}u_n(\textbf{r})
 v_n^{*}(\textbf{r})(1-2f(E_n)),\label{delta}
\end{equation}
where $g$ is the interaction constant, $f(E)$ the Fermi distribution
function and $E_c$ the cutoff energy which is related by the BCS
relation via the transition temperature $T_c$ and the
superconducting energy gap $\triangle_0$.

The vector potential $\textbf{A}(\textbf{r})$ is related to the
current distribution $\textbf{j}(\textbf{r})$ by Maxwell's equation
\begin{equation}
\nabla\times\nabla\times\textbf{A}(\textbf{r})
=\frac{4\pi}{c}\textbf{j}(\textbf{r}),\label{maxwell}
\end{equation}
where the current distribution \cite{gennes,gygi} is given by
\begin{eqnarray}
&&\textbf{j}(\textbf{r})=\frac{e\hbar}{2mi}
\sum_n\left[f(E_n)u_n^*(\textbf{r})(\nabla-\frac{ie}{\hbar
c}\textbf{A}(\textbf{r}))u_n(\textbf{r})\right.\nonumber\\
&&\left.+(1-f(E_n))v_n(\textbf{r})(\nabla-\frac{ie}{\hbar
c}\textbf{A}(\textbf{r}))v_n^*(\textbf{r})
-H.c.\right].\label{current}
\end{eqnarray}

The chemical potential $\mu$ is determined by the particle number
conservation imposed on this system \cite{kato}
\begin{eqnarray}
N=2\int\sum_n\left\{f(E_n)|u_n(\textbf{r})|^2
+(1-2f(E_n))|v_n(\textbf{r})|^2\right\},\label{number}
\end{eqnarray}
where $N$ is the total number of particles.

The boundary conditions for the above equations are given by
\begin{eqnarray}
\textbf{n}\cdot
\left(\frac{\hbar}{i}\nabla-\frac{e}{c}\textbf{A}\right)u_n=0,\ \ \
\ \textbf{n}\cdot
\left(\frac{\hbar}{i}\nabla+\frac{e}{c}\textbf{A}\right)v_n=0\label{boundary},
\end{eqnarray}
where we consider two dimensional superconducting samples placed in
the $(x, y)$ plane, which are immersed in insulating medium in the
presence of a perpendicular uniform magnetic field along $z$
direction. $\textbf{n}$ is the normal vector of the boundary.

We numerically solve Eqs.(\ref{BdG})-(\ref{boundary})
self-consistently, based on finite elements method
\cite{zienkiewicz, suematsu}. All input parameters used in the
calculation are microscopic parameters which can in principle be
obtained from band structure calculations \cite{gygi}. These
parameters consist of the cutoff energy $E_c$, the coupling constant
$g$ and $k_F\xi_0$ \cite{hayashi}, where $k_F$ ($v_F$) is the Fermi
wave number (velocity) and $\xi_0=\hbar v_F/\triangle_0$ the
coherence length.

In this paper, we consider the system at the temperature $T=0.1T_c$
and choose $E_c=5\triangle_0$ and $k_F\xi_0=2$. To conveniently
compare our results with the experiment, other parameters are chosen
so as to make the Ginzburg-Landau parameter $\kappa$ to be a
specified value, where $\kappa=0.96\lambda_L(0)/\xi_0$ \cite{gennes}
and $\lambda_L(0)$ is the London penetration depth. We consider a
square superconductor with the length of a side $a=5\xi_0$ and
$\kappa=20$. The sample is first cooled down into the
superconducting state, and then we slowly increase the magnetic
field and solve the BdG equations at each field, mimicing the zero
field cooling measurements \cite{cren,nishio2}. The global
magnetization $M$ can be calculated as
$M=1/(2S)\int_S\textbf{r}\times\textbf{j} dxdy$, where $S$ is the
sample area. In the frame of BdG theory, the free energy
$\mathcal{F}$ of the system is given \cite{kosztin} by
\begin{eqnarray}
&&\mathcal{F}=2\sum_nE_nf(E_n)-2\sum_nE_n\int
d^3\textbf{r}|v_n(\textbf{r})|^2\nonumber\\
&\!\!&-2k_BT\sum_n\{f(E_n)\ln
f(E_n)+[1-f(E_n)]\ln[1-f(E_n)]\}\nonumber\\
&&+\int d^3\textbf{r}\frac{|\triangle(\textbf{r})|^2}{g}+\int
d^3\textbf{r}\frac{|\textbf{H}(\textbf{r})-\textbf{H}_0|^2}{8\pi},\nonumber
\end{eqnarray}
where $k_B$ is the Boltzmann constant, $\textbf{H}_0$ the external
magnetic field and $\textbf{H}(\textbf{r})$ the spatial dependent
magnetic field inside the sample.

\begin{figure}[tbp]
\centering
\renewcommand{\figurename}{FIG. }
\includegraphics[width=0.42\textwidth]{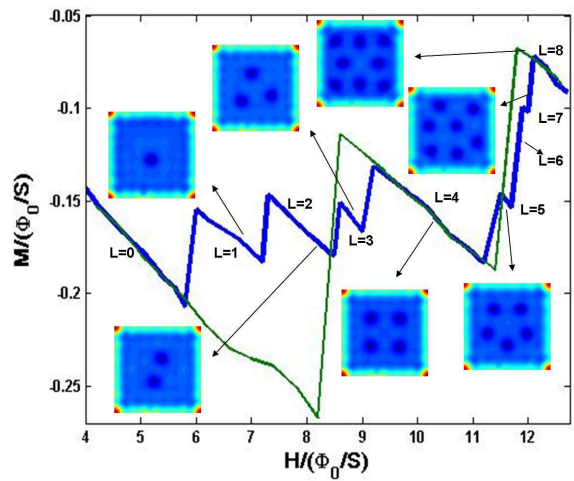}
\caption{(Color online) Magnetization (\textbf{M}) versus the
external magnetic field (\textbf{H}) for a superconducting square
sample without (thin line) and with (thick line) one defect. The
vorticity $L$ are labeled near the branches. For the sample with one
defect, the corresponding vortex pattern at each $L$ is given as the
contour plots on the insets.}
\end{figure}

Shown in Fig. 1 is the $\textbf{H}_0$ dependent magnetization of the
square sample. We can see that the magnetization shows a zigzag-like
pattern. Each descending branch of the zigzag can be described by a
fluxoid number $L$ (vorticity), which determines how many times the
phase of pair potential $\triangle(\textbf{r})$ change by $2\pi$
along the sample's circumference. Here $L$ is just the number of
vortices. Sweeping the magnetic field continuously, the
magnetization evolves along one of the fluxoid curves until it
reaches its end and jumps to the next curve, belonging to another
fluxoid state.

In a perfect square sample with four-fold symmetry, we find that
vortices enter the sample in unit of 4 with $L=4n$ ($n=0,1,2,...$),
and they replicate the geometrical symmetry of the sample, see Fig.
1. Note that the vortices enter the sample as 4 individual single
quantum vortices rather than a giant vortex, which is in agreement
with the previous works \cite{suematsu,kim}. However, if we reduce
the sample size or increase the temperature, the vortices merge into
one giant vortex. Further, upon the entrance of vortices, the
magnetization displays a sudden jump to the next branch. And
correspondingly the free energy of the system drops discontinuously.
This is consistent with the fact that the superconductor under
consideration here is a type II one, which characterizes a negative
surface energy \cite{bonca}. Besides, we also study samples with
other geometries, such as rectangle which has a two-fold symmetry,
and disk which is rotation-invariant. In the case of rectangle, our
results show that the vortices penetrate the sample in pair, just as
expected. While for a disk, the Meissner state persists until
$H>12\Phi_0/S$, indicating that the critical field for vortex
penetration is much larger than other geometry, which has been
reported in experiment \cite{geim1}. Thus, the above results suggest
that the geometrical symmetry of sample determines how the vortices
entre the sample.

When the symmetry of sample is broken by placing one defect on the
sample boundary, the situation is dramatically changed. As an
example, we consider the square sample with one sharp defect located
at $2/5$ of the length at the bottom edge, which mimics the
practical experiment realized in ref. \cite{geim1}. While in the
perfect sample, the vortices enter in unit of 4, in the sample with
one defect, the vortices enters one by one. Furthermore, from the
vortex pattern (see the inset in Fig. 1), we can see that when the
number of vortices inside the sample becomes large, the increased
interaction between vortices dominates over the influence of the
boundary, which makes the triangular lattice a favorable
arrangement. The system free energy follows the same trend as in the
perfect square, once a vortex enters the sample, the free energy of
system makes a discontinuous drop to a lower energy level. This
confirms the determining influence of the sample symmetry on the
vortex penetration and arrangement.

Now we address how the geometrical symmetry determines the vortex
entrance and which intrinsic material parameter controls the
nucleation and entrance of the vortices. We calculated the current
density inside the samples $\textbf{j}(\textbf{r})$. The current
distributions at several representative fields are shown in Fig. 2.
It is evident that the current distribution conforms to the
geometrical symmetry. In the perfect square sample (Fig. 2, contour
plot I), the current density distribution has a four-fold symmetry,
forming 4 equivalent hot spots where the current density is the
highest. Upon increasing the magnetic field, the current density at
the hot spots increases monotonously until it reaches a critical
current $j_c$ simultaneously. Further increasing the magnetic field
results in the nucleation of vortices at the hot spots, and at the
same time the current density drops to a magnitude much smaller than
$j_c$. For the sample with one defect (Fig. 2, contour plot II), the
process is the same except that there is only one hot spot, due to
the addition of the defect. Consequently, the change of current
distribution leads to the large deviation of vortex entrance
behavior. We notice that the critical current density for the vortex
nucleation $j_c\sim2.5en\hbar/m\xi_0$ is roughly equal to the
depairing current density derived from London theory,
$J_c=cH_c/4\pi\lambda=2.56en\hbar/m\xi_0$. These facts suggest that
in the mesoscopic superconductors, the intrinsic material parameter
that determines the nucleation of vortices is the depairing current
density.

\begin{figure}[tbp]
\centering
\renewcommand{\figurename}{FIG. }
\includegraphics[width=0.45\textwidth]{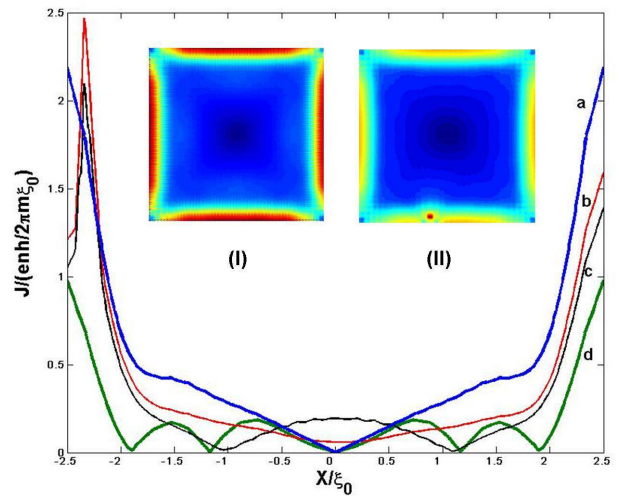}
\caption{(Color online) The spatial dependent current density of the
sample. The insets (I) and (II) are the contour plots of the
magnitude of current density for the sample without and with a
defect, respectively. The lines corresponds to the intersecting line
of the current density in (I) and (II). The solid line \textbf{a}
and \textbf{d} are for the perfect sample under the magnetic field
$H=8.3\Phi_0/S$ (\textbf{a}, before the vortex entrance) and
$H=8.4\Phi_0/S$ (\textbf{d}, after the vortex entrance); the thin
line \textbf{b} and \textbf{c} are the intersecting line across the
defect, under the magnetic field $H=5.6\Phi_0/S$ (\textbf{c}, before
the vortex entrance) and $H=5.8\Phi_0/S$ (\textbf{b}, after the
vortex entrance);}
\end{figure}

We further consider the cases where there are two defects on the
sample boundary. When the defects distribute symmetrically, for
example, the two defects are at the middle positions of opposite
edges, the sample has two fold symmetry and two equivalent hot spots
are formed around the two defects. Thus the vortices enter the
sample in pair through the defects, as shown in Fig. 3a. On the
contrary, if the defects are not equivalent, there is a difference
between the maximum current density at the defects. As a result, the
vortex penetrates the sample one by one, and the vortex nucleates
first at the defect with higher maximum current density.

Therefore, our results support such a scenario for the vortex
nucleation and entrance in the mesoscopic superconductor: When an
external magnetic field is applied, before the penetration of
vortices, the magnetic field is screened out by the circulating
current, resulting in a Meissner state. The spatial distribution of
current is determined by the symmetry of the sample geometry and the
distribution of defects. The points where the magnitude of current
are maximum will form hot spots. As the magnetic field increases,
the current density at the hot spots increases correspondingly. Once
the current density at the hot spots reach the depairing current
density, vortices will nucleate at the hot spots and enter the
superconductor. Accompanying the entrance of vortex, the system
magnetization and free energy have one discontinue drop and the
system jumps to the next quantum state through a first order phase
transition. Thus the geometric symmetry of the sample influences the
spatial distribution of the hot spots, resulting in different vortex
entrance behaviors. The influence of geometrical symmetry is a
general property of the confined vortex system, thus this scenario
can be easily generalized to the confined superfluids and
Bose-Einstein condensates.

\begin{figure}[tbp]
\centering
\renewcommand{\figurename}{FIG. }
\includegraphics[width=0.5\textwidth]{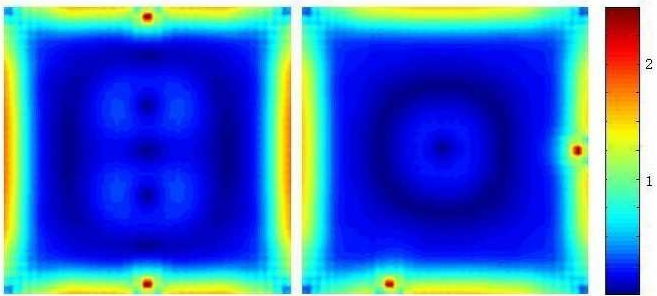}
\caption{(Color online) The contour plots of the magnitude of
current density of the sample with two defects at the magnetic field
$H=6.2\Phi_0/S$. At this magnetic field, two vortices (left) and one
vortex (right) has penetrated into the sample. The left is the
sample with two defects arranged symmetrically at the middle of up
and bottom edges, while in the right, two defects are not
equivalent.}
\end{figure}

In conclusion, by developing an effective numerical method to solve
the BdG equations self-consistently, the effect of the geometrical
symmetry on vortex penetration in mesoscopic superconductors and its
mechanism are studied quantitatively. We demonstrated that, the
condition of the nucleation of vortex is that the current density at
hot spots reaches the depairing current density. The geometrical
symmetry influences the vortex nucleation and entrance through
determining the spatial distribution of hot spots. The entrance of
vortices leads to one first-order transition between the quantum
states with different number of vortices. Our results suggest that
by modifying the topological factors such as geometrical symmetry
and defects etc., it is possible to control the entrance and
location of the vortices, and thus the quantum states. In practical
experiment, one can control the quantum states of a mesoscopic
superconductor by tuning the applied magnetic field or current
density. This opens up an applicable way to manipulate the quantum
states in mesoscopic superconductors, which is crucial for
nano-devices based on mesoscopic superconductors.

The authors would like to thank fruitful discussions with Jiang-Ping
Hu. This work was supported by NSFC under grants Nos. 10974241,
10874235, 10934010, 60978019, the NKBRSFC under grants Nos.
2009CB929100, 2009CB930701 and 2010CB922904.

\end{document}